\documentclass{article}
\usepackage{spconf,amsmath,graphicx}
\usepackage{xcolor} 
\usepackage{gensymb} 
\usepackage{multirow}

\title{On HRTF Notch Frequency Prediction\\Using Anthropometric Features and Neural Networks}
\name{\begin{tabular}{c}Lior Arbel\textnormal{\textsuperscript{1}} \kern2em Ishwarya Ananthabhotla\textnormal{\textsuperscript{2}} \kern2em Zamir Ben-Hur\textnormal{\textsuperscript{2}} \\ David Lou Alon\textnormal{\textsuperscript{2}} \kern2em Boaz Rafaely\textnormal{\textsuperscript{1}}\end{tabular}}
\address{$^1$ School of Electrical and Computer Engineering,\\Ben-Gurion University of the Negev, Beer-Sheva 84105, Israel\\$^2$ Reality Labs Research, Meta, 1 Hacker Way, Menlo Park, CA 94025, USA}
\begin{document}
%
\maketitle
\begin{abstract}
High fidelity spatial audio often performs better when produced using a personalized head-related transfer function (HRTF). However, the direct acquisition of HRTFs is cumbersome and requires specialized equipment. Thus, many personalization methods estimate HRTF features from easily obtained anthropometric features of the pinna, head, and torso. The first HRTF notch frequency (N1) is known to be a dominant feature in elevation localization, and thus a useful feature for HRTF personalization. This paper describes the prediction of N1 frequency from pinna anthropometry using a neural model. Prediction is performed separately on three databases, both simulated and measured, and then by domain mixing in-between the databases. The model successfully predicts N1 frequency for individual databases and by domain mixing between some databases. Prediction errors are better or comparable to those previously reported, showing significant improvement when acquired over a large database and with a larger output range.
\end{abstract}
\section{Introduction} \label{sec:intro}
Binaural reproduction of spatial audio often achieves best results when personalized using the listener's unique head-related transfer function (HRTF) \cite{Guezenocnew}. However, HRTF acquisition is cumbersome and requires specialized equipment \cite{andreopoulou2015inter}. A need exists for HRTF personalization methods which may be employed by end-users, not requiring specialized equipment. HRTF is a product of the pinna, head, and torso physiology \cite{pelzer2020head}. Thus, various approaches for HRTF personalization are based on prediction from anthropometric features. One approach involves direct prediction of the entire HRTF, or of distinct aspects such as magnitude and specific frequency bins \cite{bilinski2014hrtf}, which are explicitly used for reproduction. Another approach is to predict different HRTF features in order to inform HRTF selection. In this approach highly indicative HRTF features are predicted. Then, a best matching HRTF is selected from a database. Across all approaches, prediction is performed using different anthropometric feature representations, such as raw geometric features \cite{pelzer2020head}, autoencoded features \cite{chen2019autoencoding}, images \cite{zhi2022towards}, and 3D representations \cite{warnecke2022hrtf, zhou2021predictability}.

Distinct HRTF features seem to correspond to different localization tasks, thus HRTF feature prediction may be conducted towards specific performance goals. Horizontal localization is dominated by binaural cues, namely interaural time difference (ITD) and interaural level difference (ILD). These cues primarily stem from head shape and so may be straightforwardly approximated \cite{xie2013head}. On the other hand, vertical localization is dominated by monaural spectral cues originating from pinna features which are reflected in the HRTF's high frequencies \cite{iida2007median}. 

Most HRTFs comprise one or more notches in the frequency range above $5\,$kHz, termed N1, N2, et cetera. The notches' central frequencies are dominant vertical localization cues \cite{iida2014personalization}. Therefore, HRTF personalization methods focusing on vertical localization may rely on notch frequency \cite{onofrei20203d, spagnol2021estimation}.

Several studies focus on notch frequency prediction from anthropometric features \cite{onofrei20203d, iida2021estimation, miccini2019estimation}. Typically, prediction is conducted using small measured HRTF databases of 20-80 ears and geometric pinna input features. Sometimes, more elaborate 3D input features are used \cite{spagnol2021estimation}. However, obtaining these features likely requires specialized equipment, making them unsuitable for an end-user. Reported prediction errors are typically in the $0.05-0.1\,$octave range, while the just noticeable difference (JND) for N1 notch frequency is $0.1 - 0.2\,$octave \cite{iida2014personalization}. The small database size affects the prediction accuracy and limits the prediction method to multivariate linear regression. Indirectly, small databases are also likely to have a smaller output range, which in turn somewhat facilitates the prediction, reduces the estimation error, but diminishes generalizability. Existing literature does not describe N1 frequency prediction using very large databases with a sufficiently large output range, or by machine learning methods from simple pinna features.

This paper presents N1 frequency prediction from pinna geometry using three HRTF databases: CHEDAR - a large parameterized-simulated database \cite{ghorbal2020computed}, HUTUBS - a small database of both simulated and measured data \cite{brinkmann2019hutubs}, and a small, measured, internal database. The databases are larger, and comprise a larger output range, than those used in existing works. N1 frequency is predicted by a neural model for each database separately. In addition, domain mixing is employed to enhance predictions on the small databases, with the goal of enabling future listening tests. The neural model outperforms linear regression in most cases when tested over a large number of naive subjects spanning a large output range. Domain mixing further improves predictions for both HUTUBS and the internal database. The results indicate that N1 frequency may be predicted in high accuracy given sufficiently large databases, and that prediction may be performed on small databases by domain mixing, under specific conditions.

\section{Data Preparation}\label{sec:databases} 

\begin{table*}[th]
\centering
\begin{tabular}{|l|ll|ll|ll|ll|}
\hline
\textbf{Model}         & \multicolumn{2}{c|}{\textbf{CHEDAR}}                 & \multicolumn{2}{c|}{\textbf{\begin{tabular}[c]{@{}c@{}}HUTUBS\\ (simulated)\end{tabular}}} & \multicolumn{2}{c|}{\textbf{\begin{tabular}[c]{@{}c@{}}HUTUBS\\ (measured)\end{tabular}}} & \multicolumn{2}{c|}{\textbf{M1}}                     \\ \hline
                       & \multicolumn{1}{l|}{RMS Hz}          & octave        & \multicolumn{1}{l|}{RMS Hz}                             & octave                           & \multicolumn{1}{l|}{RMS Hz}                             & octave                          & \multicolumn{1}{l|}{RMS Hz}          & octave        \\ \hline
Naive prediction       & \multicolumn{1}{l|}{1151}            & 0.152         & \multicolumn{1}{l|}{1014.19}                            & 0.13                             & \multicolumn{1}{l|}{1075.1}                             & 0.16                            & \multicolumn{1}{l|}{927.2}           & 0.15          \\ \hline
Linear                 & \multicolumn{1}{l|}{953.3}           & 0.17          & \multicolumn{1}{l|}{1060.68}                            & 0.15                             & \multicolumn{1}{l|}{\textbf{1045.2}}                    & \textbf{0.17}                   & \multicolumn{1}{l|}{922.19}          & 0.15          \\ \hline
Neural                 & \multicolumn{1}{l|}{\textbf{505.07}} & \textbf{0.05} & \multicolumn{1}{l|}{1033.64}                            & 0.15                             & \multicolumn{1}{l|}{1047.86}                            & 0.17                            & \multicolumn{1}{l|}{969.82}          & 0.14          \\ \hline
Domain mixing (CHEDAR) & \multicolumn{1}{l|}{N/A}             & N/A           & \multicolumn{1}{l|}{\textbf{980.5}}                     & \textbf{0.15}                    & \multicolumn{1}{l|}{1105.23}                            & 0.175                           & \multicolumn{1}{l|}{1175.69}         & 0.19          \\ \hline
Domain mixing (HUTUBS) & \multicolumn{1}{l|}{N/A}             & N/A           & \multicolumn{1}{l|}{N/A}                                & N/A                              & \multicolumn{1}{l|}{N/A}                                & N/A                             & \multicolumn{1}{l|}{\textbf{896.99}} & \textbf{0.14} \\ \hline
\end{tabular}
\caption{N1 frequency prediction errors across databases and models. In some cases, the neural model outperforms linear prediction. Domain mixing enhances predictions when both source and target domain are acquired by the same method.}
\label{results_table}
\end{table*}

\subsection{Databases}
HRTF prediction problems pose an inherent trade-off. Data\-bases acquired from human subjects are useful for conducting listening tests. However, such databases are typically small, and limit the complexity of possible prediction models. On the other hand, large databases are typically parameterized, and thus unsuitable for listening tests. In an attempt to mitigate this trade-off, this work involves three HRTF databases: CHEDAR, HUTUBS, and an internal database, denoted as \textit{``M1''}. Prediction is performed for each database separately, and domain mixing is performed in-between the databases.

CHEDAR is a simulated parameterized database of 1253 subjects with corresponding anthropometric data. The head and torso are identical across all subjects, and the right and left ears are identical for each subject. The anthropometric data consists of standard CIPIC features: seven pinna distances ($d_1 - d_7$) and two pinna angles (rotation and flare) \cite{algazi2001cipic}.

HUTUBS is a database acquired from human subjects, consisting of simulated and measured HRIRs for both ears, and standard CIPIC anthropometric features. HUTUBS's total size is 182 ears, excluding repetitions, dummy heads and subjects without corresponding anthropometric data. M1 is an internal database consisting of measured HRIRs of both ears of 96 human subjects. In addition, M1 contains anthropometric data in the form of 53 keypoint coordinates per ear, extracted from a 3D scan of the subject's pinnas, and the two standard CIPIC pinna angles. 

For both HUTUBS and M1, both left and right ears are utilized for prediction, maximizing the databases' size. Note that mixing right and left ears is made possible due to the fact that head and torso are known not to affect notch frequency \cite{iida2018median}, and that the head and torso's effect on the HRIR is removed during the notch extraction process.

\subsection{Input Features} \label{sec_inputfeatures}
A total of 9 CIPIC standard input features are used: seven pinna distances ($d_1 - d_7$), and two pinna angles (rotation and flare). While for CHEDAR and HUTUBS all features are explicitly provided, for M1 the distances are calculated from a set of keypoint coordinates on the pinna. However, since the keypoints were not originally positioned with CIPIC features in consideration, they only approximate CIPIC distances, as described in detail in Section \ref{discussion}.

\subsection{Output}
N1 frequency was extracted from all databases using the method described in \cite{iida2014personalization}, for the front direction HRIR (azimuth $0\degree$, elevation $0\degree$) of all unique ears. The extraction method consists of clipping the HRIR by a window centered around the HRIR's maximum, and performing FFT on the result. The resulting spectrum's maximum is recorded as P1, and the lowest local minimum above P1 is recorded is N1. 

Only notch frequencies above $5\,$kHz are known to contribute to elevation localization. In addition, some subjects lack an apparent notch. Therefore, subjects with notches lower than $5\,$kHz or without a prominent notch were removed from the datasets, resulting in a total number of examples of 903 (CHEDAR), 182 (HUTUBS simulated), 171 (HUTUBS measured), and 191 (M1). The N1 frequency in each datasets ranges from $6\,$kHz to over $11\,$kHz ($\sim0.9\,$octave). For HUTUBS, a discrepancy exists between N1 frequency values extracted from measured and simulated HRIRs of each subject. The average RMS difference is $1560\,$Hz, and the correlation coefficient is $0.32$.

\section{Methods} \label{prediction}
\subsection{Single Database Prediction} \label{sec:single_prediction}
\begin{figure}[tbp]
    \centering
    \includegraphics[width=\columnwidth]{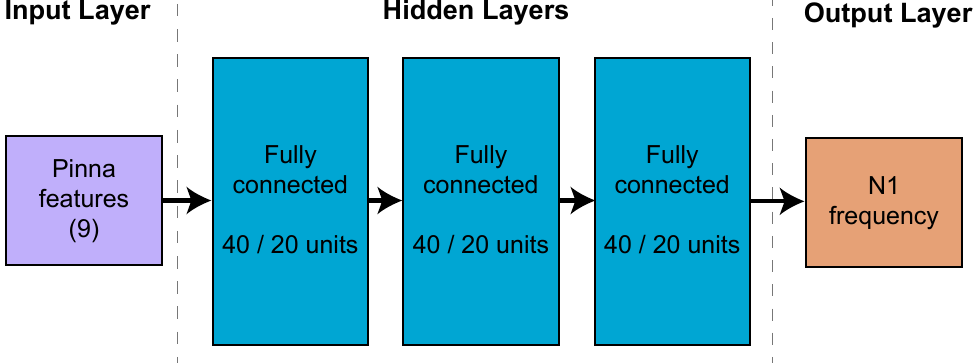} 
    \caption{The neural model used for N1 prediction, consisting of three fully connected hidden layers.}
    \label{fig:block_diagram}
\end{figure}

The N1 frequency was separately predicted for each database using both linear regression and a neural network. 
Linear regression was performed to compare against existing results in literature, and to establish a benchmark for neural model prediction. Neural model prediction was performed by a model consisting of three fully connected hidden layers of 40 units each (CHEDAR, M1) or 20 units each (HUTUBS), shown in Figure \ref{fig:block_diagram}. The model was trained using 60\% of the examples, with 20\% used as a validation set, and the remaining examples used as a test set. 

In order to asses how database size effects prediction performance, the neural model prediction was also performed with increasing training set sizes, and a fixed-size test set of 25 examples. For CHEDAR, the training set size ranged from $50-700$ examples. For HUTUBS and M1, the training set ranged from $50-150$ examples. 

\subsection{Domain Mixing} \label{shift}

Both HUTUBS and M1 are insufficiently large to obtain predictions below JND, as described in Section \ref{results}. This limitation may be mitigated by domain mixing \cite{weiss2016survey}. Several domain mixing configurations were performed: CHEDAR as source domain and HUTUBS as target domain; CHEDAR as source domain and M1 as target domain; and HUTUBS (measured) as source domain and M1 as target domain. For each configuration, the target domain training set was duplicated, with each example appearing twice, and mixed with the source domain training set. The model was then trained on the mixed training set, and tested on the target domain test set.

\section{Results} \label{results}
Errors of N1 frequency prediction for all databases and prediction methods are shown in Table \ref{results_table}. The values represent test set averages over nine executions with varied randomization. The results are also compared to naive prediction - prediction of the database's mean output as a fixed value. For CHEDAR, linear regression resulted in $953.3\,$Hz RMS error ($0.17\,$octave) and neural model prediction in $505.07\,$Hz RMS error ($0.05\,$octave). The neural model's errors are significantly lower than the lower JND threshold of $0.1\,$ octave. For CHEDAR, the neural model's predictions are comparable or better than predictions reported in literature \cite{iida2014personalization, onofrei20203d}, derived from simple pinna features, and pertain to a much larger test~set. 

For HUTUBS, comparable results are obtained over the simulated and measured HRIRs. For HUTUBS (simulated), the neural model marginally outperforms linear regression by approximately $30\,$Hz RMS , but does not outperform naive prediction. For HUTUBS (measured), linear regression and neural prediction achieve similar results, about $30\,$Hz RMS lower than naive prediction. For M1, linear regression outperforms both neural prediction and naive prediction. For both HUTUBS and M1, the overall errors are higher than the lower JND threshold. 

\begin{figure}[tbp]
    \centering
    \includegraphics[width=\columnwidth]{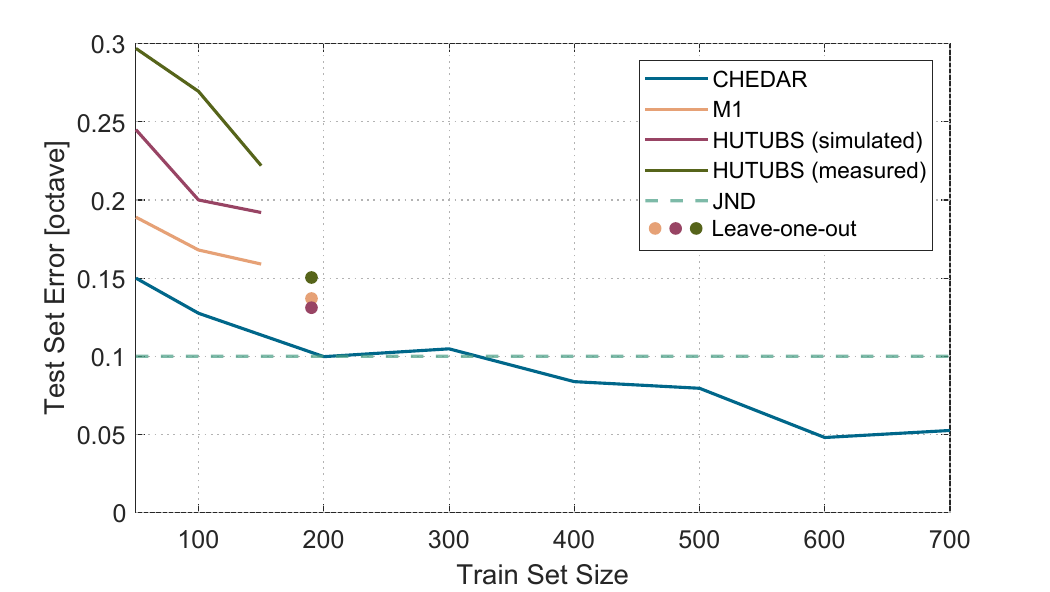} 
    \caption{N1 prediction errors versus training set size. The errors decrease with size. For CHEDAR, 200-300 examples are required to meet lower JND threshold, while M1 and HUTUBS are insufficiently large.}
    \label{fig:error_vs_train}
\end{figure}

Prediction errors versus training set size are shown in Figure \ref{fig:error_vs_train}. Prediction errors decrease as the training set size increases, across all databases. For CHEDAR, 200-300 examples are required to meet the lower JND threshold, as discussed further in Section \ref{discussion}. This size exceeds the entire small databases. Figure \ref{fig:error_vs_train} also includes additional data points for M1 and HUTUBS, which represent a $180$-example training set achieved by a leave-one-out approach. These points are not directly comparable to the rest of the data points. However, they demonstrate the potential of further increasing the training set size.

Domain mixing yielded varied results. With CHEDAR as source domain and HUTUBS (simulated) as target domain, it outperforms the neural model by approximately $50\,$Hz. However, for HUTUBS (measured) as target domain, the performance is worse than the neural model's, and for M1 it is the worst of all methods. However, domain mixing with HUTUBS (measured) as source domain and M1 as target domain outperforms all other methods.

\section{Discussion} \label{discussion}
The results obtained with CHEDAR confirm the relation between N1 frequency and pinna features, and demonstrate that this relation is non-linear. Theoretically, given mean prediction errors of these magnitudes and a database acquired on human subjects, HRTF selection and listening tests may be performed. However, accuracy with HUTUBS and M1 is significantly lower and insufficient for HRTF selection.

The performance differences may be explained by several reasons. Primarily, database size seems to be crucial. About 200-300 examples are required for sufficiently accurate N1 frequency prediction, while most measured databases are smaller. In addition, simulated databases may be easier to predict than measured databases in some cases. HRIR measurements and corresponding N1 frequency values may have inaccuracies stemming from the acquisition process. It is also possible that some simulations over-simplify the relation between pinna features and HRIR, and are therefore easier to predict. The differences between simulations and measurements are the only explanation for the performance distinction observed between HUTUBS's measured and simulated HRIRs for some training set sizes.

An additional reason may explain M1's high errors: M1's keypoint positions do not fully correspond to CIPIC distances, and are thus only approximations of CIPIC features. M1's features might be comparatively less pertinent to notch prediction than CIPIC features.

\begin{figure}[t]
    \centering
    \includegraphics[width=\columnwidth]{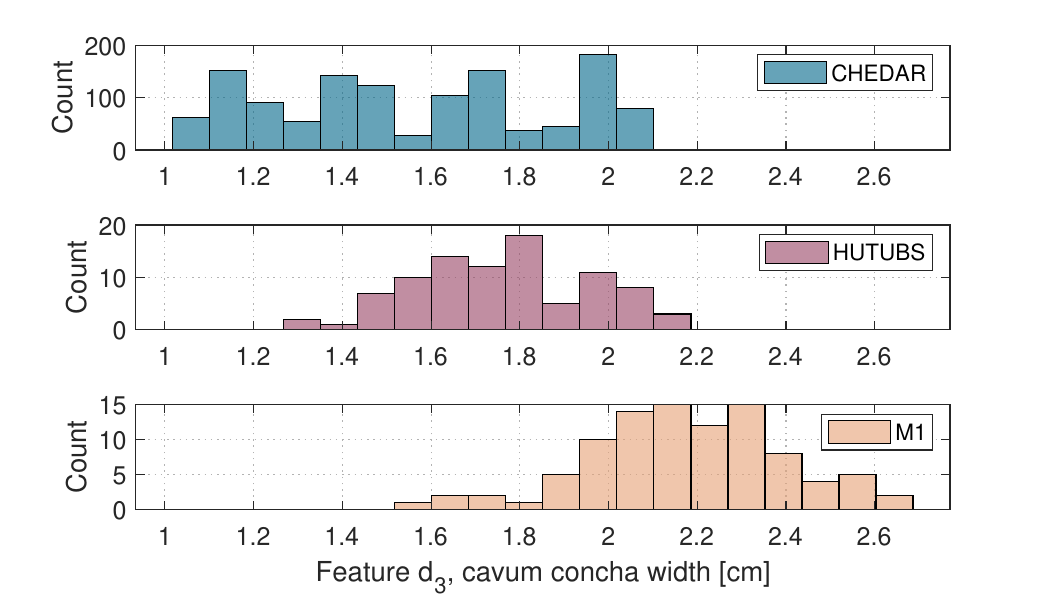} 
    \caption{Feature $d_3$ (cavum concha width) distributions for different databases. HUTUBS's distribution is partly contained within CHEDAR's, while M1's range is considerably higher. This and similar discrepancies may explain the inconsistent domain mixing results.}
    \label{fig:feature_dist}
\end{figure}

Domain mixing with CHEDAR as source domain is successful on HUTUBS (simulated), but not on M1. This may be attributed to domain shift. For all features, HUTUBS distributions are mostly contained within CHEDAR distributions. However, M1's distributions are significantly different than CHEDAR's for some features. For features $d_2, d_3$ and $d_7$, there is little to no overlap of distributions, as demonstrated in Figure \ref{fig:feature_dist}. For features $d_4, d_6$, rotation and flare, the distributions partly overlap. Only for features $d_1$ and $d_5$, the distributions considerably overlap. The domain shift may stem from the incompatible placement of keypoints in M1 as discussed above, or simply reflect the databases' population variance.   

The output acquisition method (simulations or measurements) also seems to be a key factor. When both the source and target domain were acquired by the same method, domain mixing outperforms all other methods. However, when the source and target domain are of different acquisition methods, domain mixing does not offer an advantage over other methods.

These factors may be considered when preparing data\-bases for HRTF predictions. Feature distributions of simulated databases may be selected in the first place to overlap with existing measured databases. Measured databases may be created with keypoints corresponding to standard pinna features, primarily CIPIC. It is possible that domain mixing conducted on databases meeting these requirements would achieve better results. For the purpose of listening tests on M1, other databases for domain mixing may be explored. Such databases may be simulated and with a better feature compatibility, or measured with a larger number of examples.

\section{Conclusion}
This paper presented a neural model for N1 frequency prediction from pinna features. The model was employed on three databases, consisting of both measured and simulated HRIRs. Results demonstrate that in certain configurations, notch frequency may be predicted to be below the JND given a sufficiently large database, estimated at about $200-300$ examples. Prediction errors on small databases were further reduced by domain mixing with examples of other database acquired by the same method. However, domain mixing in-between simulated and measured databases did not improve the predictions. As the model uses raw pinna distances and angles, it may be potentially employed by end users of spatial audio products to inform HRTF personalization. In this scenario, the features will be extracted from an ear photo acquired by the user. Further work should focus on successful domain mixing between measured databases, either by obtaining larger databases or by mixing additional databases. Given successful predictions on measured databases, subjective listening tests should then be performed in order to validate N1 frequency as a basis for HRTF personalization. In addition, the prediction may be expanded to include other elevations and azimuths.
\vfill\pagebreak
\bibliographystyle{IEEEbib}
\bibliography{hrtf}

\end{document}